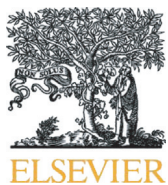
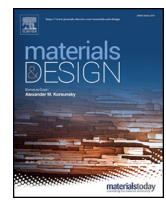

# Impact of interfaces on the radiation response and underlying defect recovery mechanisms in nanostructured Cu-Fe-Ag

Michael Wurmshuber [a,\*], David Frazer [b], Andrea Bachmaier [c], Yongqiang Wang [d], Peter Hosemann [b], Daniel Kiener [a]

[a] Department of Materials Physics, Montanuniversität Leoben, Jahnstraße 12, 8700 Leoben, Austria
[b] Department of Nuclear Engineering, University of California Berkeley, 4155 Etcheverry Hall, Berkeley, CA 94720, USA
[c] Erich-Schmid Institute of Materials Science, Austrian Academy of Sciences, Jahnstraße 12, 8700 Leoben, Austria
[d] Material Science and Technology Division, Los Alamos National Laboratory, Bikini Atoll Rd., SM 30, Los Alamos, NM 87545, USA

## HIGHLIGHTS

- A novel solid-state route was applied to fabricate Cu-Fe-Ag nanocomposites with different micro- and nanostructure.
- Irradiation of the nanostructured materials results in radiation tolerant behavior, originating in the vast amount of interfaces.
- An explanation for the observed radiation response was attempted based on accommodation and emission of defects at interfaces.

## GRAPHICAL ABSTRACT

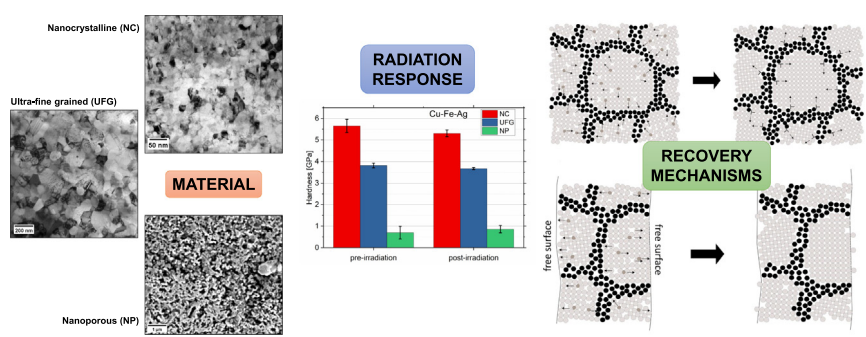

## ABSTRACT

Newest developments in nuclear fission and fusion technology as well as planned long-distance space missions demand novel materials to withstand harsh, irradiative environments. Radiation-induced hardening and embrittlement are a concern that can lead to failure of materials deployed in these applications. Here the underlying mechanisms are accommodation and clustering of lattice defects created by the incident radiation particles. Interfaces, such as free surfaces, phase and grain boundaries, are known for trapping and annihilating defects and therefore preventing these radiation-induced defects from forming clusters. In this work, differently structured nanocomposite materials based on Cu-Fe-Ag were fabricated using a novel solid-state route, combining severe plastic deformation with thermal and electrochemical treatments. The influence of different interface types and spacings on radiation effects in these materials was investigated using nanoindentation. Interface-rich bulk nanocomposites showed a slight decrease in hardness after irradiation, whereas the properties of a nanoporous material remain mostly unchanged. An explanation for this different material behavior and its link to recovery mechanisms at interfaces is attempted in this work, paving a concept towards radiation resistant materials.
© 2018 The Authors. Published by Elsevier Ltd. This is an open access article under the CC BY license (http://creativecommons.org/licenses/by/4.0/).

## 1. Introduction & motivation

The performance of materials in irradiation environments remains of great interest to the scientific community, due to numerous nuclear fission and fusion technology developments [1–4]. Moreover, the recent revival of manned spacecraft demands new structural and electronic materials to withstand high-energy solar radiation and cosmic rays over long periods of time in order to ensure safe and successful space travel in the future [5]. One of the main challenges for materials exposed to radiation is radiation-induced hardening and embrittlement. The cause for the drastic effects on mechanical properties are lattice defects generated by the incident radiation particles, creating primary knock-on ions which further cause large displacement cascades. Frenkel pairs are created by the so-called radiation damage event and subsequently form clusters such as dislocation loops, voids or stacking fault tetrahedra, which act as obstacles to dislocation movement and thus cause a harder and more brittle material behavior [1–8]. The reduction in ductility and fracture toughness is especially crucial in safety-relevant structural applications. Therefore, it is desired to create radiation tolerant materials with self-healing properties in order to ensure a longer lifetime and a reduced safety risk. Several studies have been conducted on how to suppress these radiation-induced property changes and create a material that is unaffected by irradiation. The most promising approach is the introduction of interfaces to the material that attract and annihilate the generated point defects before they can accommodate and cluster [9–28]. The most effective way to reach this goal is realized by using nanostructured materials where a large amount of interfaces are closely spaced, optimizing their effect as defect sinks.

\* Corresponding author.
E-mail address: michael.wurmshuber@unileoben.ac.at (M. Wurmshuber).

There exist a variety of different fabrication methods to process nanostructured materials, such as chemical methods (e.g. sol-gel process) [29,30], deposition (e.g. electrodeposition, physical vapor deposition) [31,32] or mechanical ball milling [33–35]. Valiev et al. [36] introduced the possibility to process bulk nanocrystalline metals by the rather convenient methods of severe plastic deformation (SPD). Especially high-pressure torsion (HPT) has proven to be an effective way to reduce the grain size of a material down to the nanometer regime. During HPT a disk-shaped specimen is placed between two anvils and torsion-strained under high pressure of several GPa [36–40].

Of lately, nanoporous materials received a lot of attention, as their high surface-to-volume ratio results in unique properties such as low specific weight, high energy absorption and excellent thermal and electrical conductivity, making them suitable for various applications such as electrocatalysts, sensors, actuators, lightweight structures, dampeners and heat exchangers [41–44]. Extensive research has also been performed in recent years to investigate the behavior of nanoporous metal foams under irradiation and it was found that the vast amount of free surface in these materials acts as perfect defect sink [13,14,17,18,20,27]. As has been demonstrated for the Cu-Fe [45] and Au-Fe system [46], fabrication of nanoporous materials can be realized by selective etching of a composite material consisting of two or more metals with a large difference in electrochemical potential. For certain material combinations, the etching process has to be conducted with an applied protective potential (potentiostatic dealloying) to ensure only dissolution of one component and protection of the desired components [47–52].

The goal of this work is to fabricate an ultra-fine grained, a nanocrystalline and a nanoporous material and compare radiation-induced mechanical property changes. The obtained results are used to discuss potential recovery mechanisms at the nanoscale and unravel the influence of interface type and spacing on the radiation response.

## 2. Experimental

### 2.1. Material selection

The material surrogate system used in this work to test the hypothesis of enhanced radiation tolerance was selected to be Cu-Fe-Ag. This material system is deployed mainly in electrical applications. While copper-rich Cu-Fe-Ag alloys are utilized when a combination of high-strength and high conductivity are required [53–56], iron-rich representatives of this system are known to exhibit the giant magnetoresistance effect [57–59]. However, the main reason this system was chosen for this work was to provide a material consisting of elements which are not soluble in each other at equilibrium conditions, resulting in the creation of phase boundaries. Additionally Fe is known to be dissolved by HCl, while Cu and Ag should stay intact and form a nanoporous structure with additional Cu/Ag interfaces within the foam ligaments [60].

### 2.2. Sample fabrication

For this study, copper powder (99.9% purity, 20 μm particle size), iron powder (99% purity, 74 μm particle size) and silver powder (99.9% purity, 25 μm particle size, all powders provided by Alfa Aesar, Thermo Fisher Scientific GmbH, Karlsruhe, Germany) were mixed together at a ratio of 50 at.% Cu, 25 at.% Fe and 25 at.% Ag. The mixture was compacted using hot-isostatic pressing (HIP) at 630 °C under a pressure of 30 MPa for 15 min under vacuum (Fig. 1a)). These parameters were chosen based on experience with processing similar materials. The resulting material cylinders were cut into 1 mm thick discs with a radius of 8 mm and subsequently deformed to an equivalent Von Mises strain of 1450 using a high-pressure torsion (HPT) tool (Fig. 1b)) [40]. The SPD applied to the material results in a supersaturated single-phase of the normally immiscible components [61,62], as was also observed for ball-milled material samples in the Cu-Fe-Ag system [58,59]. To allow a decomposition of the phases, two different heat treatments were performed in a muffle furnace (K1252, Heraeus GmbH, Hanau, Germany; Fig. 1c)). A heat treatment at 400 °C for 1 h yielded a nanocrystalline (NC) polyphased material (grain size <100 nm), whereas a heat treatment at 600 °C for 1 h after deformation allows enough grain growth for the resulting material to be classified as ultra-fine grained (UFG; grain size 100–500 nm). Additionally, several UFG samples underwent potentiostatic dealloying in 2 molaric (5 wt%) HCl for 4 h at room temperature using a potentiostat (Gamry PCI4, Gamry Instruments Inc., Warminster, USA; Fig. 1d)). The protective potential for this procedure was chosen as −450 mV. This step ensures selective electrochemical dissolution of the Fe phase only, leaving behind a foam-like nanoporous (NP) material, containing a vast amount of free surface and additional phase interfaces within the Cu-Ag foam ligaments.

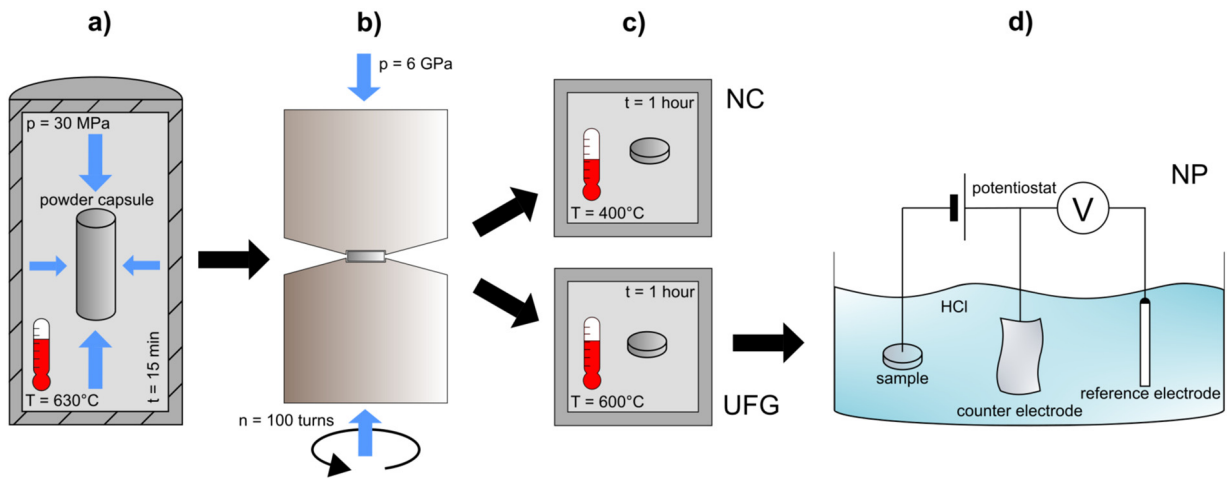

**Fig. 1.** Schematic workflow of the sample fabrication process. a) The powder mixture is compacted using hot-isostatic pressing (HIP) and b) subsequently deformed via high-pressure torsion (HPT). c) The deformed specimens are annealed at 400 °C (NC material) and 600 °C (UFG material). d) NP samples are fabricated using potentiostatic dealloying of UFG material.

The resulting microstructures of the three differently structured materials were characterized using a field emission scanning electron microscope (SEM; LEO type 1525, Carl Zeiss GmbH, Oberkochen, Germany) and a dual beam Focused Ion Beam (FIB)-SEM (Quanta 3D FEG, FEI, Hillsboro, USA). UFG and NC samples were additionally investigated using a transmission electron microscope (TEM; CM12, Philips, Amsterdam, Netherlands). The material composition was characterized via energy dispersive X-ray spectroscopy (EDX; 7426, Oxford Instruments plc, Abingdon, UK; Software EDAX and AZtec).

### 2.3. Ion beam irradiation

To study radiation effects on the nanostructured materials, samples (NC, UFG and NP) were irradiated with 1 MeV protons (hydrogen, $H^+$) at the Ion Beam Materials Laboratory (IBML) at Los Alamos National Laboratory (LANL). The beam current was 3 μA in average. The temperature was monitored with a thermocouple next to the sample and maintained below 50 °C [63,64]. The computer software "Stopping Range of Ions in Matter" (SRIM) [65] was used to simulate the irradiation and yielded a penetration depth of approximately 8 μm in the bulk samples and 10 μm in the NP samples (assuming total dissolution of Fe). For the simulation the default displacement energy value of 25 eV was used. Protons were chosen as radiation particle because of their relatively large penetration depth (compared to heavy ions), their high dose rate (compared to neutrons) and the fact that they do not activate the material, making proton-irradiation a safe and convenient state-of-the-art method for radiation damage and effect studies [66–68]. Additionally, protons account for a large portion of space radiation and can be found in e.g. galactic cosmic radiation, solar winds or solar particle events (so-called "proton storms") [5,69]. Therefore the proton irradiation-response of possible spacecraft materials is of great interest. The materials were irradiated to a nominal dose of 1 dpa in the plateau regime before the stopping peak, which yielded about 13 dpa at the stopping peak (10 dpa in NP material). The corresponding dose profiles are depicted in Fig. 2.

### 2.4. Nanoindentation experiments

The limited penetration depth of the proton irradiation calls for small-scale testing methods to correctly assess radiation-induced mechanical property changes without unintended probing of unirradiated material volume [67,68]. As the need of small-scale methods arises, nanoindentation seems to be an excellent choice, as only little sample preparation is required to gain insight into the mechanical behavior of the material.

Indentation experiments were conducted at room temperature using a diamond Berkovich tip. All used tips were calibrated on fused silica to obtain a correct area function and ensure an accurate analysis of indentation experiments. For a comprehensive radiation effect study, it is crucial to use the indenter in displacement-controlled mode, to always sample the same material volume and thus dose range [67].

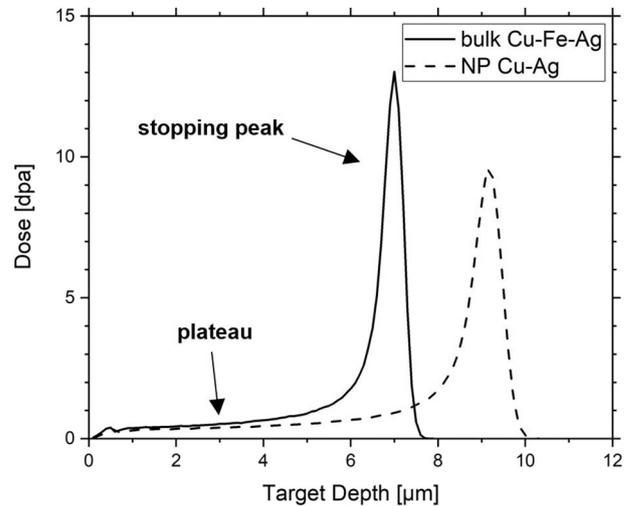

**Fig. 2.** Dose profile for 1 MeV proton irradiation of bulk (UFG and NC) Cu-Fe-Ag and NP Cu-Ag materials.

Indentation of the unirradiated materials was performed on a Micro Materials Nanotest Platform 3 nanoindenter (Micro Materials Inc., Wrexham, UK). UFG and NC samples were indented to a depth of 1000 nm. Due to the higher surface roughness and porosity, NP samples were tested to a depth of 1000 and 2000 nm. This ensures that the indents are considerably larger than the pores and that the results represent the average hardness and not individual filaments or local variations in composition and structure. The load was applied with a fixed rate of 5 mN/s, a dwell period of 15 s at the peak load and an unloading rate of 10 mN/s. To take into account thermal drift, a thermal drift correction was conducted after each indentation, using the last 60% (36 s) of the recorded drift data. The post-irradiation tests were performed using a Keysight Nano Indenter G200 (Keysight Technologies Inc., Santa Rosa, USA). To obtain as much information as possible on the material response after irradiation, continuous stiffness measurement (CSM) indentation experiments were performed parallel to the ion-beam to a displacement of 2500 nm as well as cross-sectional to a displacement of 400 nm. All indents were analyzed following the method of Oliver and Pharr [70].

## 3. Results

### 3.1. Microstructure

The microstructural evolution of UFG and NC samples is depicted in Fig. 3a)–d). One can distinguish the different phases by their phase contrast, i.e. their ability to scatter electrons. The lightest element is therefore appearing darkest in SEM images recorded with a backscattered electron detector and vice versa (Fig. 3a)). The metastable single-phase that was present after HPT deformation (Fig. 3b)) successfully separated into its three original components after heat treating at 400 °C (NC material) and 600 °C (UFG material), as is also apparent from the selected area diffraction (SAD) patterns (insets in the corresponding TEM images). From TEM images one can use the grain intercept method to calculate the grain size in NC material to be 18.8 (±1.8) nm in average, in contrast to an average grain size of 95.7 (±10.3) nm in the UFG material (Fig. 3c) and d)). Consequently, the grain size of the UFG material in this study is actually at the boundary between UFG and NC regime (~100 nm). However, for easier differentiation of the two materials, the term UFG was chosen for this nanostructured material.

Potentiostatic dealloying of the UFG material yielded a nanoporous foam (Fig. 3e)). After 4 h of dealloying total dissolution of iron could not be achieved (as is apparent from the measured chemical composition in

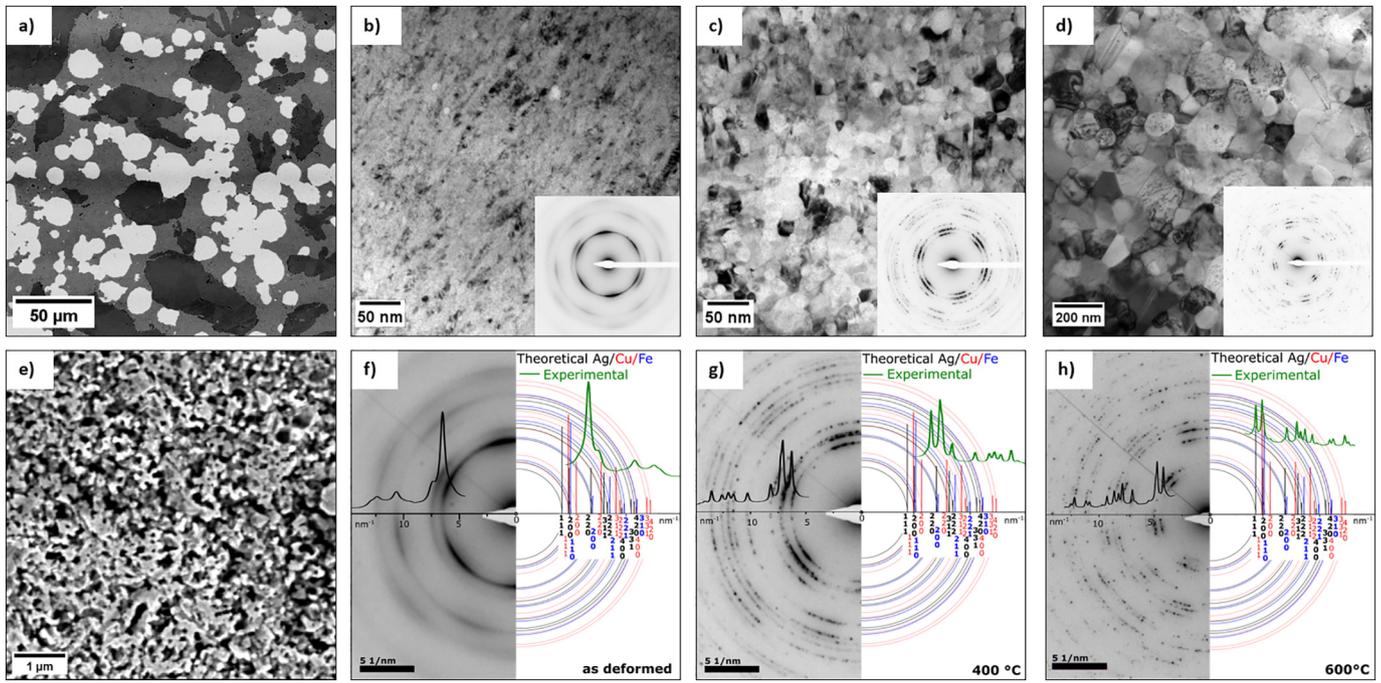

**Fig. 3.** Microstructural evolution of all samples. a) SEM image of the as-HIPed material. b) Bright field TEM image of the as-deformed material after HPT. A few oxides are visible. c) Bright field TEM image of the NC material sample. d) Bright field TEM image of the UFG material sample. e) SEM image of the NP foam sample. SAD patterns from the TEM investigations of f) as-deformed material, g) NC material and h) UFG material are analyzed and compared to the theoretical peaks of Cu, Fe and Ag.

Table 1) and the porosity of the foam showed a high degree of inhomogeneity. However, as longer dealloying experiments caused problems with surface segregation and the general goal of fabricating a multiphased foam was accomplished, the sample that underwent the electrochemical treatment for 4 h was chosen as NP sample for the radiation effect study in this work.

The SAD patterns of the TEM investigations were analyzed by azimuthal integration and the resulting peaks were compared to the theoretical peaks of pure copper, iron and silver. It is evident that the integrated diffraction pattern of the as-deformed material (Fig. 3f)) shows no similarities with the theoretical data, indicating intermixing of the three components and a solid solution. Annealing at 400 °C (Fig. 3g)) leads to beginning phase separation as is apparent from the growing number of peaks matching with reference data. Annealing at 600 °C (Fig. 3h)) results in nearly perfect agreement of experimental and theoretical peaks and therefore complete separation of the three phases.

Fig. 4 depicts SEM images of the microstructure in unirradiated and irradiated (at the stopping peak) areas in UFG and NC material. The grain size distribution seems unaltered and no indication of radiation-induced grain growth was detected.

### 3.2. Bulk perpendicular nanoindentation

Fig. 5 shows the results of CSM-nanoindentation conducted directly onto the irradiated surface of UFG and NC material in comparison to static 1000 nm indents into the unirradiated material (black squares).

**Table 1**
Measured chemical compositions for different material states gained by EDX.

| Element | Initial mixture | | UFG sample | | NP sample | | | |
|---|---|---|---|---|---|---|---|---|
| | at.% | wt% | at.% | wt% | at.% | ± | wt% | ± |
| Cu | 50.0 | 43.7 | 48.5 | 42.0 | 47.0 | 6.0 | 37.7 | 6.9 |
| Fe | 25.0 | 19.2 | 25.2 | 19.2 | 14.9 | 4.4 | 10.5 | 3.7 |
| Ag | 25.0 | 37.1 | 26.3 | 38.7 | 38.1 | 6.3 | 51.9 | 4.7 |
| Σ | 100.0 | 100.0 | 100.0 | 100.0 | 100.0 | | 100.0 | |

All values lie within a reasonable range to the results of unirradiated material. As the stopping peak (the region where most displacement damage is done) of the irradiation lies at a depth of approximately 7 μm and the plastic zone probed by nanoindentation is known to be 5–10 times bigger than the indentation depth [67,71,72], one would expect to see the influence of the dose profile and stopping peak in the hardness over depth curves resulting in a generally higher hardness (see insets in Fig. 5). As this is not the case, it can be concluded that both materials do not exhibit a pronounced hardening effect after proton irradiation. However, performing indentation on the irradiated surface yields a fairly complicated situation in terms of analyzing mechanical properties, as several superposing effects (e.g. dose profile, indentation size effect and other surface effects) can occur [67]. In order to shed more light on the material response after irradiation, additional cross-sectional indents were performed.

### 3.3. Bulk cross-sectional nanoindentation

To get a better correlation between dose and hardness, cross-sectional indents were performed on the UFG and NC material in irradiated as well as unirradiated areas perpendicular to the ion beam direction as proposed earlier [67,68] (insets in Fig. 6).

Representative nanoindentation results for UFG samples are depicted in Fig. 6(a), with the black curves representing data from the unirradiated material and the dashed red curves representing results of the indents performed at the stopping peak (approximately 7 μm from the surface). It is apparent that the hardness in irradiated areas is slightly lower than the reference indents in unirradiated areas. The average hardness (analyzed between an indentation depth of 300 and 400 nm for all valid indentation experiments) was found to be approximately 0.15 ± 0.12 GPa or 3.8% lower than in unirradiated areas.

In the NC material (Fig. 6(b)) the difference in average hardness between irradiated and unirradiated regions is observed to be even more pronounced, resulting in a hardness reduction of approximately 0.34 ± 0.33 GPa or 5.9%.

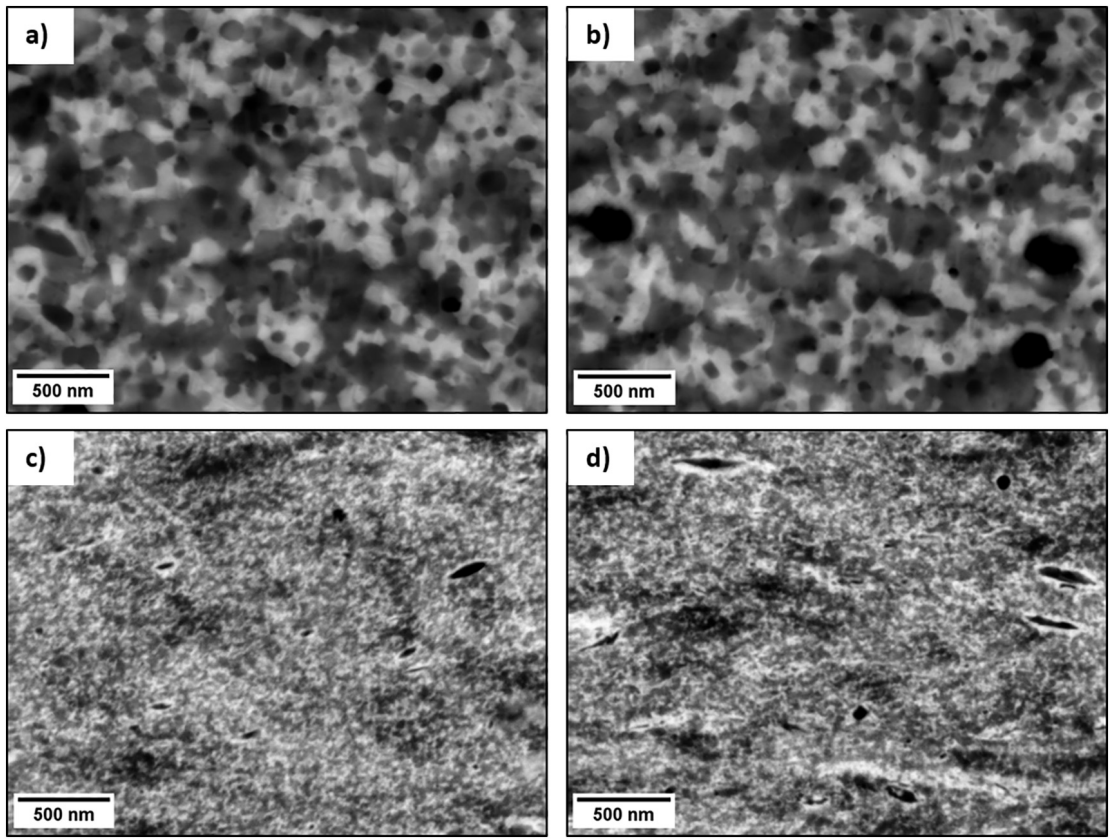

**Fig. 4.** SEM images of a) unirradiated and b) irradiated microstructure of UFG material as well as c) unirradiated and d) irradiated microstructure of NC material. No observable microstructural changes are detected. The images in irradiated state were taken from the stopping peak (~13 dpa).

### 3.4. Nanoporous nanoindentation

As mechanical or electrochemical polishing of the NP material would destroy the unique structure of the foam, cross-sectional nanoindentation is not possible. Therefore, all indents were performed directly on the irradiated surface. Due to the fact that not all of the iron was dissolved during the dealloying process, the porosity (and therefore the chemical composition) of the foam showed inhomogeneity. As every indent depends heavily on the porosity underneath, the values gained by indentation of both unirradiated and irradiated NP samples scatter, as is apparent in Fig. 7(a). Consequently, extracting only the radiation-induced changes in hardness is not a straightforward task to do. In the following, an attempt to accomplish this was made by applying the Gibson-Ashby equations [73] to estimate the porosity under each indent:

$$\sigma_y^* = C_1 \cdot \sigma_{y,s} \cdot \left(\frac{\rho^*}{\rho_s}\right)^m \quad (1)$$

$$E^* = C_2 \cdot E_s \cdot \left(\frac{\rho^*}{\rho_s}\right)^n \quad (2)$$

Here $\sigma_{y,s}$ represents the yield strength, $E_s$ the Young's modulus and $\rho_s$ the density of the solid bulk material and $\sigma_y^*$ the yield strength, $E^*$ the

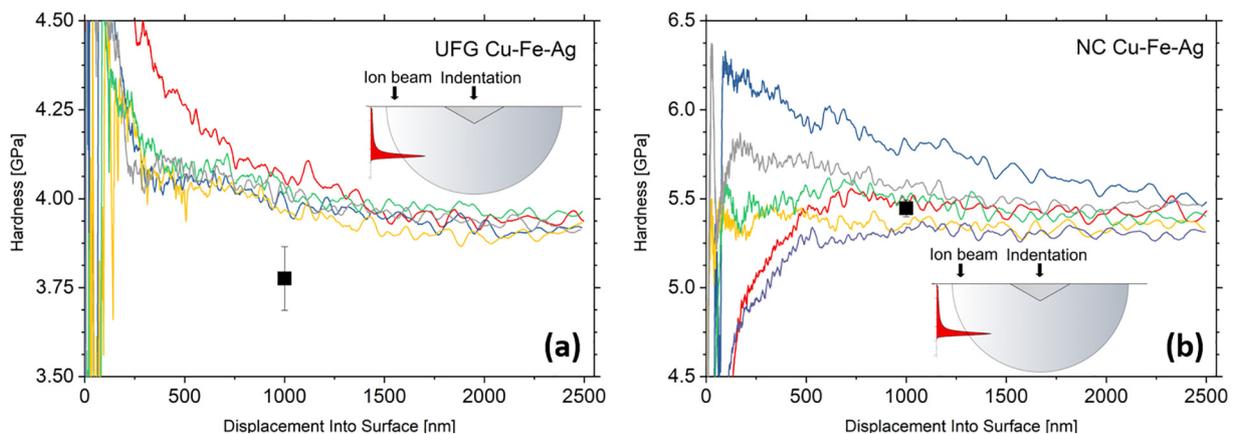

**Fig. 5.** Hardness results of nanoindentation directly onto the irradiated surface for (a) UFG and (b) NC material. Black square symbols represent static reference indents performed prior to irradiation to 1000 nm depth. The inset depicts the indent placement and probed radiation dose.

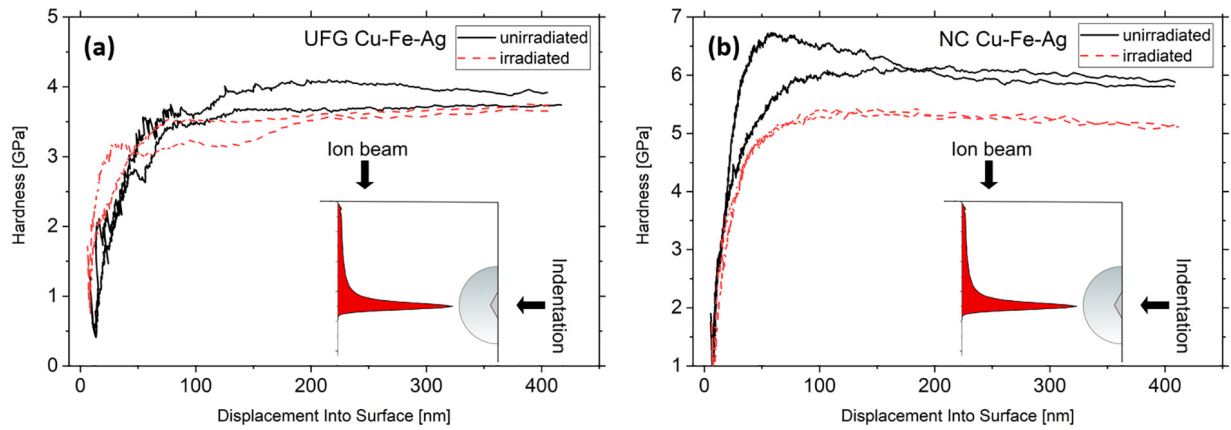

**Fig. 6.** Representative hardness results for cross-sectional nanoindentation of (a) UFG and (b) NC samples. The dashed red curves represent indents performed at the stopping peak, the black curves are reference indents conducted in the unirradiated bulk volume. The inset depicts the indent placement and probed radiation dose.

Young's modulus and $\rho^*$ the density of the porous foam. The constants in these equations describe the cell geometry and deformation behavior and are well estimated for most foams with $C_1 = 0.3$, $C_2 = 1$, $m = 1.5$ and $n = 2$. While these equations were developed for macroscopic foams and their validity for microscopic foams and nanoporous materials is still a highly discussed topic [74–77], they can provide a first estimate of mechanical properties of NP materials. Since the influence of irradiation on Young's modulus of a material is negligible, the measured modulus and the modulus of bulk UFG samples are used to estimate the density ratio $\frac{\rho^*}{\rho_s}$, i.e. porosity, under each indent using Eq. (2). As hardness and yield strength of a material are linearly correlated, this porosity can then be used in Eq. (1) to recalculate the hardness of each measurement. This was done for measurements of irradiated as well as unirradiated NP samples to help extract information about the direct influence of radiation on the mechanical properties of such a foam material. Fig. 7(b) shows the hardness over indentation depth curves gained by this approach. Although this model can only be used as a first approximation, a drastic reduction of the deviation in the data was achieved (from approximately ±1.3 GPa to ±0.3 GPa).

To provide a compact overview, the average radiation-induced hardness changes of all investigated materials are summarized in Fig. 8. It is evident that only minor changes in hardness occurred, with slight softening for the NC & UFG material upon irradiation while the NP material showed a slight increase in hardness, which is negligible considering the large error bars.

## 4. Discussion

### 4.1. UFG & NC material

The fabrication route for the UFG and NC bulk materials, including HIP, HPT and heat treatment, showed promising results and good reproducibility, making it an easy and effective way to process multiphase nanostructured bulk materials.

While nanoindentation performed directly onto the irradiated surface of UFG and NC materials might sample a variety of superposing effects, a pronounced influence of the radiation dose profile can already be ruled out when comparing pre- and post-irradiation indents in Fig. 5. This indicates a certain radiation tolerance of these materials. However, the additional cross-sectional indents in Fig. 6 shed more light on the radiation effects on mechanical properties and will be discussed in more detail in the following.

The slightly lower hardness level of irradiated UFG and NC material is contrary to the usually observed radiation hardening in conventional structural materials. Radiation-induced softening was observed before in different materials [78–81] and is usually attributed to changes in composition or structure of the materials. Jiao et al. [79] suggested radiation-induced grain growth and stress relaxations on grain boundaries as explanation for a softening effect in irradiated nanocrystalline ZrN films. The materials investigated in this work, however, are free of any observed segregation and no pronounced grain growth was detected subsequent to the irradiation (see Fig. 4; a growth of about

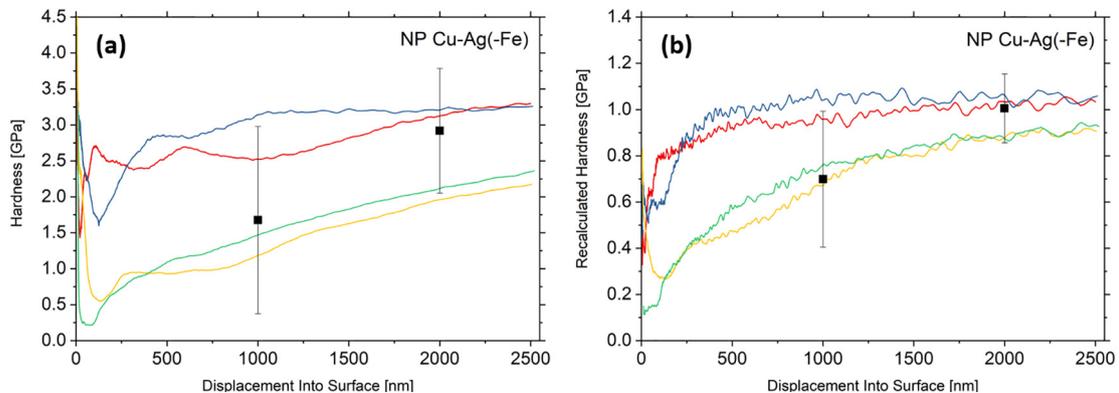

**Fig. 7.** Hardness over indentation depth results for NP material (a) before and (b) after applying a model to account for local inhomogeneity in porosity. The black square symbols represent results from static indentation before the irradiation treatment.

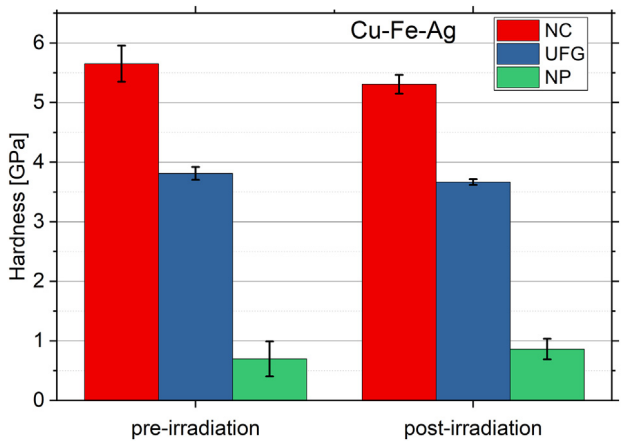

**Fig. 8.** Summarized hardness changes for all investigated materials due to proton-irradiation. NC and UFG results were analyzed at an indentation depth of 300–400 nm, NP results at a depth of 1000 nm.

20 to 50% would be necessary to account for the observed softening). Thus, a different explanation of the observed softening effect in UFG and NC materials is required and attempted in the following.

First, it should be stated that such fine-grained materials are known to contain only a few dislocations due to the hardening-by-annealing effect [82,83]. The heat treatment after the HPT deformation enhances dislocation movement and the high amount of high-angle grain boundaries present in these materials act as sinks for dislocation. The consequences are nearly dislocation-free grains and an increased hardness. As neither chemical composition nor grain size or shape was found to change after irradiation, and there are almost no dislocations in the grains to recover, it is clear that the measured softening of mechanical properties must be originated in the vast amount of interfaces in these nanostructured materials.

When the grain size approaches a regime where conventional Frank-Read sources do not fit in the grain interior anymore or would result in too high Orowan stresses, a different dislocation nucleation mechanism prevails. This mechanism is based on grain boundary (GB) distortions, where atoms associated with the GB are slightly displaced from their assigned position, forming so-called ledges (Fig. 9). These GB ledges lead to localized stress concentrations when an external force is applied. Once this increased stress intensity reaches a certain threshold value a partial dislocation is emitted from the ledge to relax the structure [20,84–91]. This partial dislocation propagates rather fast through the obstacle-free grain and is then absorbed by another GB. Irradiating a nanostructured material will introduce vacancies and interstitials to the material lattice. The majority of these radiation-induced defects will migrate to GBs and other interfaces due to the close interface-spacing in these materials. These defects are believed to distort the GBs and form additional ledges, which results in more stress concentrations and thus easier dislocation nucleation and a softer material behavior (see Fig. 10).

While in NC materials most defects are believed to migrate to interfaces, leaving only a few defects in the grain interior, the larger interface-spacing in UFG material will result in more defects remaining in the grain. These defects can form small dislocation loops or voids and act as obstacles to dislocation movement, counteracting the softening effect from facilitated dislocation nucleation from ledges. This explains the less pronounced softening in UFG material compared to NC material seen in Fig. 8. Additionally, the UFG GBs are closer to an equilibrium structure than the GBs in the less annealed NC material. Non-equilibrium GBs, generated in SPD processes, have been observed to show higher sink strengths than their equilibrium counterparts [25,27].

### 4.2. NP material

Fabrication of the NP foam material was not a straightforward task, since free corrosion in HCl, as has already successfully been demonstrated for Cu-Fe [45] and Au-Fe [46] precursors, resulted in coarsely porous, nearly pure Ag foams due to a galvanic corrosion effect. Therefore, a protective potential had to be applied during the etching process to ensure only dissolution of Fe. Out of the conditions tested within the scope of this work, the UFG precursor that underwent the potentiostatic dealloying at −450 mV for 4 h showed the most promising results. However, by adjusting parameters such as dealloying potential, dealloying time, etching solution or temperature, still better results could be achieved. This, however, is rather time-consuming and was not in the focus of this work. The hardness results for irradiated and un-irradiated NP material depend heavily on the local porosity underneath the indent, yet they all lie within the same range. The values after the applied Gibson-Ashby correction still scatter too much to claim that the material is completely unaffected by radiation. However, it can be assumed that the majority of radiation-induced defects are annihilated in this material, as free surfaces are an even more efficient defect sink than the interfaces present in UFG and NC material (see Fig. 11). This is a remarkable finding, since the foam ligament diameters in this work are considerably larger than in earlier studies about radiation resistant NP materials [13,14,17,18,20,27]. Therefore, the phase boundaries within the NP ligaments might have an additional positive effect on radiation tolerance and a complex interaction of the effects of annihilation at free surface, radiation hardening due to remaining defects in the ligaments and the radiation softening mechanisms explained above cannot be excluded based on the data gained in this work.

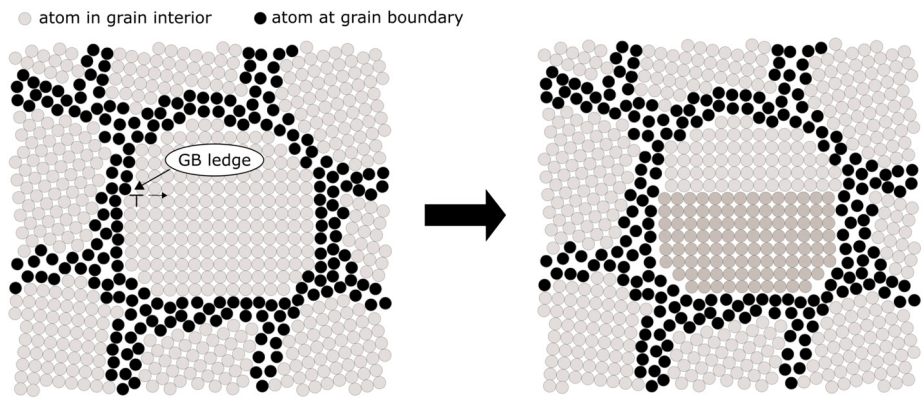

**Fig. 9.** Illustration of plastic deformation in nanocrystalline materials. GB ledges result in highly localized stress concentrations which in turn promote the nucleation of partial dislocations. The result is a stacking fault in the grain interior and a more relaxed grain boundary structure. A second partial dislocation can recover the initial atomic positions again.

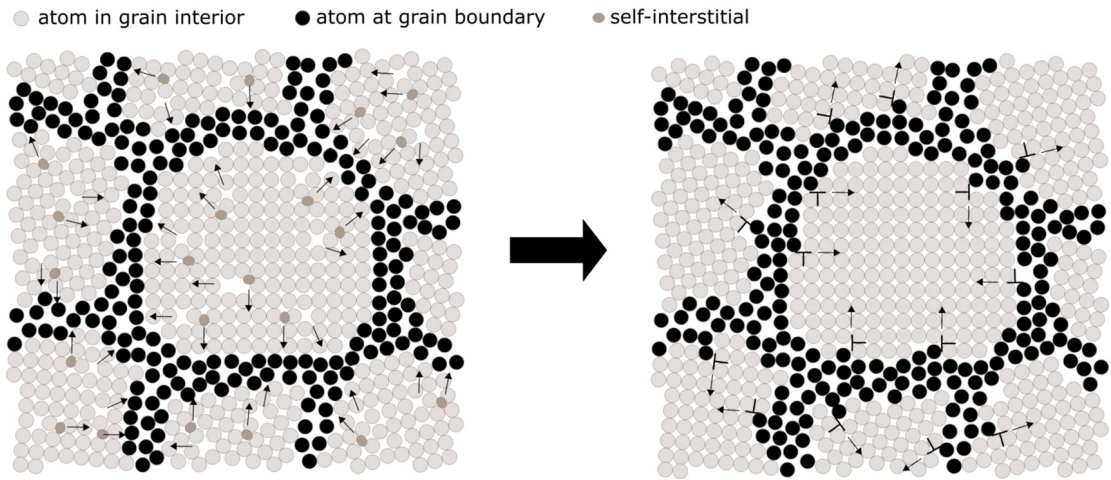

**Fig. 10.** Suggested defect recovery mechanisms in nanocrystalline material. The radiation-induced vacancies and self-interstitials migrate to the GB, where they distort the GB structure and form additional ledges. These serve as easier dislocation nucleation spots and cause a softer material behavior.

## 5. Summary & conclusion

Differently structured Cu-Fe-Ag bulk nanocomposites were fabricated using a solid-state route. The individual components were bulk mechanically alloyed via HPT and subsequently heat treated to reach phase separation and adjust the grain size. Material batches with a grain size of about 20 nm (NC material) and 100 nm (UFG material) were produced. A NP foam material was subsequently created by potentiostatic dealloying of the UFG material. Mechanical properties of the different nanostructured materials were characterized before and after an irradiation treatment with 1 MeV protons to 1 dpa of damage at near room temperature. UFG and NC material experienced a slight decrease in hardness after irradiation, which can be attributed to the interplay of defect annihilation at phase- and grain boundaries and accompanied distortions in the interface structure, resulting in facilitated dislocation nucleation. Mechanical properties of the NP material scattered markedly due to an inhomogeneity in porosity and composition. The variation in the indentation results could be reduced by accounting for the local porosity under each indent using the measured Young's modulus. No notable difference in hardness between pre- and post-irradiation state was observed before or after the applied correction, which indicates that the majority of radiation-induced defects in this material are annihilated at the vast amount of surface and additional interfaces within the foam ligaments, rendering the material radiation resistant.

In conclusion, all investigated nanostructured materials show highly radiation tolerant properties. While annihilation of radiation-induced point defects on free surface leads to complete recovery of mechanical properties, migration to and accommodation at grain and phase boundaries results in enhanced plasticity and therefore a slight decrease in hardness after irradiation. This proof-of-principle is an important first step towards novel materials for application in irradiative environments. Studies on and improvement of other application-relevant properties of these materials, such as creep, corrosion or fabrication on a large scale, are still to come and will strengthen the prospect of deploying nanostructured materials in nuclear and space travel applications.

### CRediT authorship contribution statement

**Michael Wurmshuber:** Formal analysis, Writing - original draft, Writing - review & editing, Investigation, Methodology. **David Frazer:** Methodology, Writing - review & editing, Investigation. **Andrea Bachmaier:** Resources, Writing - review & editing, Methodology. **Yongqiang Wang:**

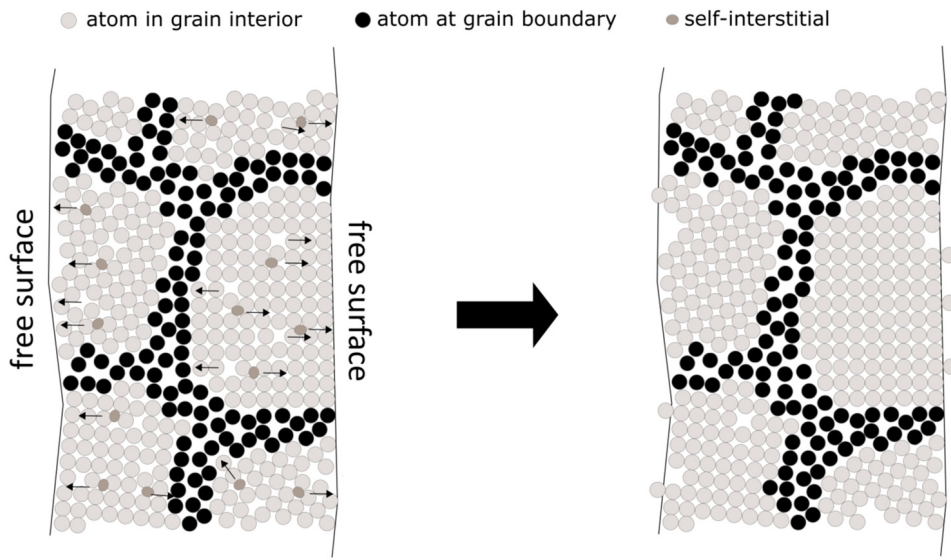

**Fig. 11.** Suggested defect recovery mechanisms in nanoporous materials. The majority of radiation-induced vacancies and self-interstitials migrates to the preferable defect sink, i.e. the free surface. There they are annihilated and the initial pre-irradiation state is almost fully recovered.

Resources, Writing - review & editing. **Peter Hosemann:** Supervision, Writing - review & editing, Resources. **Daniel Kiener:** Supervision, Writing - review & editing, Formal analysis.

### Acknowledgements

The financial support by the Austrian Marshall Plan Foundation and the University of Leoben is gratefully acknowledged. The authors also acknowledge funding by the European Research Council under Grant numbers: 771146 (MW, DK) and 757333 (AB). Instrument access was provided in part through the Biomolecular Nanotechnology Center (BNC) at UC Berkeley. The authors gratefully acknowledge the financial support under the scope of the COMET program (A2.12) within the K2 Center "Integrated Computational Material, Process and Product Engineering (IC-MPPE)" (Project No 859480). This program is supported by the Austrian Federal Ministries for Transport, Innovation and Technology (BMVIT) and for Digital and Economic Affairs (BMDW), represented by the Austrian Research Funding Association (FFG), and the Federal States of Styria, Upper Austria and Tyrol.

### Data availability

The datasets generated during the current study are available from the corresponding author on reasonable request.

### Author contributions

M.W. conducted sample fabrication, nanoindentation experiments and data analysis and wrote the main body of the paper. D.F. helped with methodology and conducting nanoindentation experiments. A.B. helped with sample fabrication and provided the TEM images. Y.W. provided beamtime for the proton irradiation. P.H. conducted the proton irradiation and co-supervised the work. D.K. supervised the work and helped with interpretation of the results. All authors contributed to the final version of the paper.

### References

[1] S.J. Zinkle, Advanced materials for fusion technology, Fusion Eng. Des. 74 (2005) 31–40.
[2] S.J. Zinkle, Fusion materials science: overview of challenges and recent progress, Phys. Plasmas 12 (2005), 58101.
[3] P. Yvon, F. Carre, Structural materials challenges for advanced reactor systems, J. Nucl. Mater. 31 (2009) 217–222.
[4] S.J. Zinkle, G.S. Was, Materials challenges in nuclear energy, Acta Mater. 61 (2013) 735–758.
[5] L.S. Novikov, V.N. Mileev, E.N. Voronina, L.I. Galanina, A.A. Makletsov, V.V. Sinolits, Radiation effects on spacecraft materials, J. Surf. Invest. 3 (2) (2009) 199–214.
[6] G.J. Dienes, Radiation effects in solids, Annu. Rev. Nucl. Part. Sci. 2 (1953) 187–220.
[7] G. Kinchin, R. Pease, The displacement of atoms in solids by radiation, Rep. Prog. Phys. 18 (1) (1955).
[8] B.D. Wirth, How does radiation damage materials? Science 318 (2007) 923.
[9] A. Misra, M.J. Demkowicz, X. Zhang, R.G. Hoagland, The radiation damage tolerance of ultra-high strength nanolayered composites, JOM 59 (2007) 62–65.
[10] M.J. Demkowicz, R.G. Hoagland, J.P. Hirth, Interface structure and radiation damage resistance in Cu-Nb multilayer nanocomposites, Phys. Rev. Lett. 100 (13) (2008), 136102.
[11] S. Wurster, R. Pippan, Nanostructured metals under irradiation, Scr. Mater. 60 (12) (2009) 1083–1087.
[12] X.-M. Bai, A.F. Voter, R.G. Hoagland, M. Nastasi, B.P. Uberuaga, Efficient annealing of radiation damage near grain boundaries via interstitial emission, Science 327 (2010) 1631–1634.
[13] E.M. Bringa, J.D. Monk, A. Caro, A. Misra, L.A. Zepeda-Ruiz, M. Duchaineau, F. Abraham, et al., Are nanoporous materials radiation resistant? Nano Lett. 12 (2012) 3351–3355.
[14] E.G. Fu, M. Caro, L.A. Zepeda-Ruiz, Y.Q. Wang, K. Baldwin, E.M. Bringa, M. Nastasi, et al., Surface effects on the radiation response of nanoporous Au foams, Appl. Phys. Lett. 101 (2012), 191607.
[15] W.Z. Han, M.J. Demkowicz, E.G. Fu, Y.Q. Wang, A. Misra, Effect of grain boundary character on sink efficiency, Acta Mater. 60 (18) (2012) 6341–6351.
[16] W.Z. Han, M.J. Demkowicz, N.A. Mara, E.G. Fu, S. Sinha, A.D. Rollett, Y.Q. Wang, et al., Design of radiation tolerant materials via interface engineering, Adv. Mater. 25 (2013) 6975–6979.
[17] I.J. Beyerlein, A. Caro, M.J. Demkowicz, N.A. Mara, A. Misra, B.P. Uberuaga, Radiation damage tolerant nanomaterials, Mater. Today 16 (2013) 443–449.
[18] C. Sun, D. Bufford, Y. Chen, M.A. Kirk, Y.Q. Wang, M. Li, H. Wang, et al., In-situ study of defect migration kinetics in nanoporous Ag with enhanced radiation tolerance, Sci. Rep. 4 (2014) 3737.
[19] B.P. Uberuaga, L.J. Vernon, E. Martinez, A.F. Voter, The relationship between grain boundary structure, defect mobility, and grain boundary sink efficiency, Sci. Rep. 5 (2015) 9095.
[20] I.J. Beyerlein, M.J. Demkowicz, A. Misra, B.P. Uberuaga, Defect-interface interactions, Prog. Mater. Sci. 74 (2015) 125–210.
[21] Y. Chen, K.Y. Yu, Y. Liu, S. Shao, H. Wang, M.A. Kirk, J. Wang, et al., Damage-tolerant nanotwinned metals with nanovoids under radiation environments, Nat. Commun. 6 (2015) 7036.
[22] J. Li, K.Y. Yu, Y. Chen, M. Song, H. Wang, M.A. Kirk, M. Li, et al., In situ study of defect migration kinetics and self-healing of twin boundaries in heavy ion irradiated nanotwinned metals, Nano Lett. 15 (5) (2015) 2922–2927.
[23] C. Sun, S. Zheng, C.C. Wei, Y. Wu, L. Shao, Y. Yang, K.T. Hartwig, et al., Superior radiation-resistant nanoengineered austenitic 304L stainless steel for applications in extreme radiation environments, Sci. Rep. 5 (2015) 7801.
[24] O. El-Atwani, J.A. Hinks, G. Greaves, J.P. Allain, S.A. Maloy, Grain size threshold for enhanced irradiation resistance in nanocrystalline and ultrafine tungsten, Mater. Res. Lett. 5 (5) (2017) 343–349.
[25] G.A. Vetterick, J. Gruber, P.K. Suri, J.K. Baldwin, M.A. Kirk, P. Baldo, Y.Q. Wang, et al., Achieving radiation tolerance through non-equilibrium grain boundary structures, Sci. Rep. 7 (1) (2017), 12275.
[26] S. Jiao, Y. Kulkarni, Radiation tolerance of nanotwinned metals – an atomistic perspective, Comput. Mater. Sci. 142 (2018) 290–296.
[27] X. Zhang, K. Hattar, Y. Chen, L. Shao, J. Li, C. Sun, K. Yu, et al., Radiation damage in nanostructured materials, Prog. Mater. Sci. 96 (2018) 217–321.
[28] J. Li, C. Fan, Q. Li, H. Wang, X. Zhang, In situ studies on irradiation resistance of nanoporous Au through temperature-jump tests, Acta Mater. 143 (2018) 30–42.
[29] C. Terrier, J.P. Chatelon, J.A. Roger, Electrical and optical properties of Sb:SnO$_2$ thin films obtained by the sol-gel method, Thin Solid Films 295 (1997) 95–100.
[30] Y. Li, M.K. Ghantasala, K. Galatsis, W. Wlodarski, Oxygen gas sensing and microstructure characterization of sol-gel prepared MoO$_3$-TiO$_2$ thin films, Proceedings of SPIE - The International Society of Optical Engineering, 3892, 1999, pp. 364–371.
[31] H. Li, F. Ebrahimi, Synthesis and characterization of electrodeposited nanocrystalline nickel-iron alloys, Mater. Sci. Eng. A 347 (2003) 93–101.
[32] P.H. Mayrhofer, C. Mitterer, L. Hultman, H. Clemens, Microstructural design of hard coatings, Prog. Mater. Sci. 51 (8) (2006) 1032–1114.
[33] J. Eckert, J.C. Holzer, C.E. Krill, W.L. Johnson, Structural and thermodynamic properties of nanocrystalline fcc metals prepared by mechanical attrition, J. Mater. Res. 7 (1992) 1751–1761.
[34] C.C. Koch, The synthesis and structure of nanocrystalline materials produced by mechanical attrition: a review, Nanostruct. Mater. 2 (1993) 109–129.
[35] C.C. Koch, Synthesis of nanostructured materials by mechanical milling: problems and opportunities, Nanostruct. Mater. 9 (1997) 13–22.
[36] R.Z. Valiev, R.K. Islamgaliev, I.V. Alexandrov, Bulk nanostructured materials from severe plastic deformation, Prog. Mater. Sci. 45 (2000) 103–189.
[37] N. Tsuji, Y. Saito, S.-H. Lee, Y. Minamino, ARB (accumulative roll-bonding) and other new techniques to produce bulk ultrafine grained materials, Adv. Eng. Mater. 5 (2003) 338–344.
[38] V. Viswanathan, T. Laha, K. Balani, A. Agarwal, S. Seal, Challenges and advances in nanocomposite processing techniques, Mater. Sci. Eng. R 54 (2006) 121–285.
[39] R. Pippan, S. Scheriau, A. Taylor, M. Hafok, A. Hohenwarter, A. Bachmaier, Saturation of fragmentation during severe plastic deformation, Annu. Rev. Mater. Res. 40 (1) (2010) 319–343.
[40] R. Pippan, S. Scheriau, A. Hohenwarter, M. Hafok, Advantages and limitations of HPT: a review, Mater. Sci. Forum 584–586 (2008) 16–21.
[41] P.S. Liu, K.M. Liang, Review functional materials of porous metals made by P/M, electroplating and some other techniques, J. Mater. Sci. 36 (2001) 5059–5072.
[42] D. Kramer, R.N. Viswanathan, J. Weissmüller, Surface-stress induced macroscopic bending of nanoporous gold cantilevers, Nano Lett. 4 (2004) 793–796.
[43] K. Bonroy, J.-M. Friedt, F. Frederix, W. Laureyn, S. Langerock, A. Campitelli, M. Sára, et al., Realization and characterization of porous gold for increased protein coverage on acoustic sensors, Anal. Chem. 76 (2004) 4299–4306.
[44] Y. Ding, M. Chen, J. Erlebacher, Metallic mesoporous nanocomposites for electrocatalysis, J. Am. Chem. Soc. 126 (2004) 6876–6877.
[45] M. Kreuzeder, M.-D. Abad, M.-M. Primorac, P. Hosemann, V. Maier, M. Rebelo de Figueiredo, D. Kiener, Fabrication and thermo-mechanical behavior of nanoporous copper, J. Mater. Sci. 50 (2015) 634–643.
[46] A. Leitner, V. Maier-Kiener, J. Jeong, M.-D. Abad, P. Hosemann, S.H. Oh, D. Kiener, Interface dominated mechanical properties of ultra-fine grained and nanoporous Au at elevated temperature, Acta Mater. 121 (2016) 104–116.
[47] J. Erlebacher, M.J. Aziz, A. Karma, N. Dimitrov, K. Sieradzki, Evolution of nanoporosity in dealloying, Nature 410 (6827) (2001) 450–453.
[48] J. Erlebacher, An atomistic description of dealloying, J. Electrochem. Soc. 151 (10) (2004) C614.
[49] J. Snyder, P. Asanithi, A.B. Dalton, J. Erlebacher, Stabilized Nanoporous metals by dealloying ternary alloy precursors, Adv. Mater. 20 (24) (2008) 4883–4886.
[50] J. Weissmüller, R.C. Newman, H.-J. Jin, A.M. Hodge, J.W. Kysar, Nanoporous metals by alloy corrosion: formation and mechanical properties, MRS Bull. 34 (08) (2009) 577–586.
[51] O. Okman, D. Lee, J.W. Kysar, Fabrication of crack-free nanoporous gold blanket thin films by potentiostatic dealloying, Scr. Mater. 63 (2010) 1005–1008.

[52] M. Graf, B. Roschning, J. Weissmüller, Nanoporous gold by alloy corrosion: method-structure-property relationships, J. Electrochem. Soc. 164 (2017) C194–C200.
[53] J.S. Song, S.I. Hong, H.S. Kim, Heavily drawn Cu–Fe–Ag and Cu–Fe–Cr microcomposites, J. Mater. Process. Technol. 113 (1–3) (2001) 610–616.
[54] H. Gao, J. Wang, D. Shu, B. Sun, Effect of Ag on the microstructure and properties of Cu-Fe in situ composites, Scr. Mater. 53 (10) (2005) 1105–1109.
[55] J.S. Song, S.I. Hong, Y.G. Park, Deformation processing and strength/conductivity properties of Cu–Fe–Ag microcomposites, J. Alloys Compd. 388 (1) (2005) 69–74.
[56] B. Sun, H. Gao, J. Wang, D. Shu, Strength of deformation processed Cu–Fe–Ag in situ composites, Mater. Lett. 61 (4–5) (2007) 1002–1006.
[57] J. Wang, P. Xiong, G. Xiao, Investigation of giant magnetoresistance in magnetic, concentrated, nanostructured alloys, Phys. Rev. B 47 (1993) 8341–8344.
[58] N.S. Cohen, E. Ahlswede, J.D. Wicks, Q.A. Pankhurst, Investigation of the ternary phase diagram of mechanically alloyed FeCuAg, J. Phys. Condens. Matter 9 (1997) 3259–3276.
[59] Q.A. Pankhurst, N.S. Cohen, M. Odlyha, Thermal analysis of metastable Fe-Cu-Ag prepared by mechanical alloying, J. Phys. Condens. Matter 10 (1998) 1665–1676.
[60] P.C. Oliveira, M. Cabral, F. Charters Taborda, F. Margarido, C.A. Nogueira, Leaching studies for metals recovery from printed circuit boards scrap, Electronics & Battery Recycling '09, 2009.
[61] A. Bachmaier, M. Kerber, D. Setman, R. Pippan, The formation of supersaturated solid solutions in Fe-Cu alloys deformed by high-pressure torsion, Acta Mater. 60 (3) (2012) 860–871.
[62] K.S. Kormout, R. Pippan, A. Bachmaier, Deformation-induced supersaturation in immiscible material systems during high-pressure torsion, Adv. Eng. Mater. 19 (4) (2017), 1600675.
[63] M.J. Demkowicz, O. Anderoglu, X. Zhang, A. Misra, The influence of $\sum 3$ twin boundaries on the formation of radiation-induced defect clusters in nanotwinned Cu, J. Mater. Res. 26 (14) (2011) 1666–1675.
[64] O. Anderoglu, M.J. Zhou, J. Zhang, Y.Q. Wang, S.A. Maloy, J.K. Baldwin, A. Misra, He$^+$ ion irradiation response of Fe–TiO$_2$ multilayers, J. Nucl. Mater. 435 (1–3) (2013) 96–101.
[65] J.F. Ziegler, J.P. Biersack, U. Littmark, The Stopping Range of Ions in Matter, Pergamon Press, New York, 1985.
[66] G.S. Was, J.T. Busby, T. Allen, E.A. Kenik, A. Jenssen, A.M. Bruemmer, J. Gan, et al., Emulation of neutron irradiation effects with protons: validation of principle, J. Nucl. Mater. 300 (2002) 198–216.
[67] D. Kiener, A.M. Minor, O. Anderoglu, Y.Q. Wang, S.A. Maloy, P. Hosemann, Application of small-scale testing for investigation of ion-beam-irradiated materials, J. Mater. Res. 27 (2012) 2724–2736.
[68] P. Hosemann, C. Shin, D. Kiener, Small scale mechanical testing of irradiated materials, J. Mater. Res. 30 (2015) 1–15.
[69] E.R. Benton, E.V. Benton, Space radiation dosimetry in low-earth orbit and beyond, Nucl. Inst. Methods Phys. Res. B 184 (2001) 255–294.
[70] W.C. Oliver, G.M. Pharr, An improved technique for determining hardness and elastic modulus using load and displacement sensing indentation experiments, J. Mater. Res. 7 (1992) 1564–1583.
[71] K.L. Johnson, The correlation of indentation experiments, J. Mech. Phys. Solids 18 (2) (1970) 115–126.
[72] D. Kiener, R. Pippan, C. Motz, H. Kreuzer, Microstructural evolution of the deformed volume beneath microindents in tungsten and copper, Acta Mater. 54 (10) (2006) 2801–2811.
[73] L.J. Gibson, M.F. Ashby, Cellular Solids: Structure and Properties, 11th ed. Cambridge University Press, Cambridge, 1997.
[74] A.M. Hodge, J. Biener, J.R. Hayes, P.M. Bythrow, C.A. Volkert, A.V. Hamza, Scaling equation for yield strength of nanoporous open-cell foams, Acta Mater. 55 (2007) 1343–1349.
[75] D. Farkas, A. Caro, E.M. Bringa, D.A. Crowson, Mechanical response of nanoporous gold, Acta Mater. 61 (2013) 3249–3256.
[76] N.J. Briot, T. Kennerknecht, C. Eberl, T.J. Balk, Mechanical properties of bulk single crystalline nanoporous gold investigated by millimetre-scale tension and compression testing, Philos. Mag. 94 (2014) 847–866.
[77] N. Mameka, K. Wang, J. Markmann, E.T. Lilleodden, J. Weissmüller, Nanoporous gold - testing macro-scale samples to probe small-scale mechanical behavior, Mater. Res. Lett. 4 (2015) 27–36.
[78] A. Sato, T. Mifune, M. Meshii, Irradiation softening in pure iron single crystals, Phys. Status Solidi 18 (1973) 699–709.
[79] L. Jiao, K.Y. Yu, D. Chen, C. Jacob, L. Shao, X. Zhang, H. Wang, Radiation tolerant nanocrystalline ZrN films under high dose heavy-ion irradiations, J. Appl. Phys. 117 (2015), 145901.
[80] A. Tsepelev, Radiation-induced softening of Fe-Mo alloy under high-temperature electron irradiation, IOP Conference Series: Materials Science and Engineering, 130, 2016, p. 12016.
[81] A. Tsepelev, A.S. Ilyushin, T.Y. Kiseleva, E.A. Brovkina, V.N. Melnikov, Radiation-induced changes in the structure and mechanical properties of Fe-Mo alloy under electron irradiation, Inorg. Mater. Appl. Res. 8 (2017) 378–381.
[82] X. Huang, N. Hansen, N. Tsuji, Hardening by annealing and softening by deformation in nanostructured metals, Science 312 (2006) 249–251.
[83] O. Renk, A. Hohenwarter, K. Eder, K.S. Kormout, J.M. Cairney, R. Pippan, Increasing the strength of nanocrystalline steels by annealing: is segregation necessary? Scr. Mater. 95 (2015) 27–30.
[84] M.A. Meyers, A. Mishra, D.J. Benson, Mechanical properties of nanocrystalline materials, Prog. Mater. Sci. 51 (4) (2006) 427–556.
[85] H. Van Swygenhoven, J.R. Weertman, Deformation in nanocrystalline metals, Mater. Today 9 (2006) 24–31.
[86] H. Van Swygenhoven, Footprints of plastic deformation in nanocrystalline metals, Mater. Sci. Eng. A 483–484 (2008) 33–39.
[87] A. Hunter, I.J. Beyerlein, Unprecedented grain size effect on stacking fault width, APL Mater. 1 (3) (2013), 32109.
[88] A. Hunter, I.J. Beyerlein, Stacking fault emission from grain boundaries: material dependencies and grain size effects, Mater. Sci. Eng. A 600 (2014) 200–210.
[89] I.A. Ovid'ko, A.G. Sheinerman, R.Z. Valiev, Dislocation emission from deformation-distorted grain boundaries in ultrafine-grained materials, Scr. Mater. 76 (2014) 45–48.
[90] L.E. Murr, Dislocation ledge sources: dispelling the myth of frank–read source importance, Metall. Mater. Trans. A 47 (12) (2016) 5811–5826.
[91] V. Turlo, T.J. Rupert, Grain boundary complexions and the strength of nanocrystalline metals: dislocation emission and propagation, Acta Mater. 151 (2018) 100–111.